\documentclass[8pt,prl,twocolumn,showpacs]{revtex4}
\usepackage{amsfonts}
\usepackage{amsmath}
\usepackage{amssymb}
\usepackage{graphicx}

\providecommand{\U}[1]{\protect\rule{.1in}{.1in}}
\providecommand{\U}[1]{\protect\rule{.1in}{.1in}}
\providecommand{\U}[1]{\protect\rule{.1in}{.1in}}
\providecommand{\U}[1]{\protect\rule{.1in}{.1in}}

\ifx\pdfoutput\relax\let\pdfoutput=\undefined\fi
\newcount\msipdfoutput
\ifx\pdfoutput\undefined\else
\ifcase\pdfoutput\else
\ifx\paperwidth\undefined\else
\ifdim\paperheight=0pt\relax\else\pdfpageheight\paperheight\fi
\ifdim\paperwidth=0pt\relax\else\pdfpagewidth\paperwidth\fi
\fi\fi\fi
\ifx\pdfoutput\relax\let\pdfoutput=\undefined\fi
\newcount\msipdfoutput
\ifx\pdfoutput\undefined\else
\ifcase\pdfoutput\else
\ifx\paperwidth\undefined\else
\ifdim\paperheight=0pt\relax\else\pdfpageheight\paperheight\fi
\ifdim\paperwidth=0pt\relax\else\pdfpagewidth\paperwidth\fi
\fi\fi\fi

\begin{document}

\title{Buffering plasmons in nanoparticle waveguides at the
virtual-localized transition.}
\author{Ra\'{u}l A. Bustos-Mar\'{u}n$^{1,2}$, Eduardo A. Coronado$^{2}$ and
Horacio M. Pastawski$^{1}$}

\begin{abstract}
We study the plasmonic energy transfer from a locally excited nanoparticle
(LE-NP) to a linear array of small NPs and we obtain the parametric
dependence of the response function. An analytical expression allows us to
distinguish the extended resonant states and the localized ones, as well as
an elusive regime of virtual states. This last appears when the resonance
width collapses and before it becomes a localized state. Contrary to common
wisdom, the highest excitation transfer does not occur when the system has a
well defined extended resonant state but just at the virtual-localized
transition, where the main plasmonic modes have eigenfrequencies at the
passband edge. The slow group velocity at this critical frequency enables
the excitation buffering and hence favors a strong signal inside the chain.
A similar situation should appear in many other physical systems. The
extreme sensitivity of this transition to the waveguide and LE-NP parameters
provides new tools for plasmonics.
\end{abstract}

\affiliation{$^{1}$IFEG and FAMAF, UNC, Ciudad Universitaria, 5000
C\'{o}rdoba, Argentina.}
\affiliation{$^{2}$INFIQC, Departamento de
Fisicoqu\'{\i}mica, Facultad Ciencias Qu\'{\i}micas, UNC, Ciudad
Universitaria, 5000 C\'{o}rdoba, Argentina.}

\maketitle

\section{I. INTRODUCTION}

Electromagnetic energy can be focused and guided below the light diffraction
limit by transforming it into a collective plasmonic excitation, the surface
plasmon polariton, that can propagate along a one dimensional array of
nanoparticles (NPs) \cite{libros}. This feature has attracted significant
attention due to its potential applications in optoelectronic devices,
sub-wavelength waveguides, random lasers, optical traps, and hot-spot based
plasmonic sensors for ultrasensitive spectroscopy \cite%
{libros,Brongersma,Ejem-Util}. Previous works have already addressed the
question of how to achieve a high degree of localization of plasmonic
excitations \cite{Excitation} and have studied the plasmon propagation on
the waveguide formed by an ordered NP chain \cite{Disp-Chain,Brongersma}.
However, a fundamental question remains open: how to transform a \textit{%
localized} excitation into a strong signal somewhere else inside a finite
chain. A natural idea, would be to exploit the divergent local density of
states of 1-D systems, \textit{i.e.} the vanishing group velocity at the
band edge, as proposed for light buffering in photonic waveguides \cite%
{Baba-photonic}. However, such divergences are a bulk property absent in
finite and semi-infinite chains, where the local density of states near the
extremes cancels out at the passband edge. In a quantum tight-binding model,
this corresponds to a semi-elliptical density of states. Therefore, the only
alternative would seem to tailor the surface inhomogeneity to generate
ad-hoc resonances \cite{Newns-SantosSchmickler} that could easily be excited
and transfer energy. This poses a big challenge for both experiments and
numerical simulations, since there is a broad range of parameters to be
explored. Thus, this work resorts to a model that, containing only
essentials, could be solved analytically by using a response function
formalism. Besides the expected \textit{resonant} and \textit{localized}
eigenmodes, we find an elusive regime of \textit{virtual} states. This
appears when the resonance width collapses and before becoming a localized
state, providing for a continuity between them. Quite surprisingly, we prove
that \textit{virtual} to \textit{localized} states transition provides the
route to optimal excitation transfer by recovering, at the chain extremes, a
divergent local density of states with slow group velocity. The work is
organized as follow: In Section II we present the model used to describe the
system and develop a response function formalism and a pole analysis for it.
Then, in Section III we show and discuss the results, and finally, in
Section IV we summarize the main conclusions of the work.
\begin{figure}[th]
\begin{center}
\includegraphics[trim=0.5in 0.3in 0.3in 0.0in, width=2.7in]{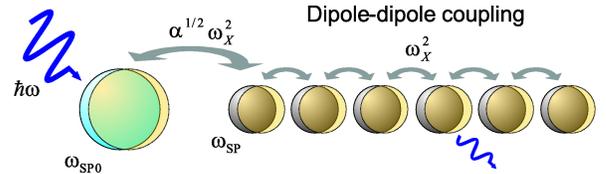}
\end{center}
\caption{(Color online) - An external source ($\hbar \protect\omega $)
excites the surface plasmon ($\protect\omega _{\mathrm{SP}0}$) of a NP which
is coupled, through $\protect\alpha ^{1/2}\protect\omega _{\text{\textrm{X}}%
}^{2}$, to a NP waveguide, of bandwidth $|\protect\omega ^{2}-\protect\omega %
_{\mathrm{SP}}^{2}|\leq 2\protect\omega _{\text{\textrm{X}}}^{2}$, where
detection takes place.}
\label{Figura-sistema}
\end{figure}

\section{II. LINEAR RESPONSE IN NANOPARTICLE WAVEGUIDES}

The system studied, see Fig. \ref{Figura-sistema}, is a linear array of $N$
metal NPs coupled to a locally excited (LE) NP, described in the coupled
dipole approximation \cite{Disp-Chain,Brongersma,CDA,Sarychev}. The induced
dipole moment $\overrightarrow{P_{i}}$ at $i^{\mathrm{th}}$ NP satisfies:%
\begin{align}
\left[ \omega _{\mathrm{SP}i}^{2}-\omega ^{2}-\mathrm{i}\eta _{i}\omega %
\right] \overrightarrow{P}_{i}& =\tfrac{1}{3}r_{i}^{3}\omega _{\mathrm{P}%
i}^{2}4\pi \epsilon _{0}  \label{PvsE} \\
& \left[ \overrightarrow{E}_{i}^{(\mathrm{ext})}+\overset{N}{\underset{j\neq
i}{\sum }}\overrightarrow{E}_{j,i}(\overrightarrow{P}_{j},\overrightarrow{d}%
_{j,i},\vec{k})\right] .  \notag
\end{align}%
Here, $r_{i}$, $\omega _{\mathrm{P}i}$, $\omega _{\mathrm{SP}i}$ and $\eta
_{i}$ correspond to the radius, bulk and surface plasmon frequencies, and
electronic damping. $\epsilon _{0}$ is the free space permittivity. $%
\overrightarrow{E}_{i}^{(\mathrm{ext})}$ and $\overrightarrow{E}_{j,i}$ are
respectively the external field and the electric field at the $i^{th}$ site
produced by the $j^{th}$ NP. In general, $\overrightarrow{E}_{j,i}$ is a
complex function that depends on the separation vector $\overrightarrow{d}%
_{j,i}=d_{j,i}\hat{d}_{j,i}$ between NPs and the wave vector $\vec{k}$.
However, if $d$ is small, $\overrightarrow{E}_{j,i}$ can be evaluated in the
near field approximation:%
\begin{equation}
\overrightarrow{E}_{j,i}(\overrightarrow{P}_{j},\overrightarrow{d}_{j,i},%
\vec{k})\overset{kd\rightarrow 0}{\approx }\frac{\overrightarrow{P}_{j}-3%
\widehat{d}_{j,i}(\overrightarrow{P}_{j}\cdot \widehat{d}_{j,i})}{4\pi
\epsilon _{0}n^{2}d_{j,i}^{3}},  \label{Enear2}
\end{equation}%
where $n$ is the refractive index of the host material. For a linear array
of NPs, plasmon oscillations can only be transverse\textbf{\ }($T$) or
longitudinal ($L$) to the chain axis, and due to the cubic dependence of $E$
on $d$, it is a good approximation to neglect contributions beyond nearest
neighbors \cite{Brongersma}. Arranging all $\overrightarrow{P}_{i}$ and $%
\overrightarrow{E}_{i}^{(\mathrm{ext})}$as vectors $\mathbf{P}$ and $\mathbf{%
E}$\textbf{,} Eq. \ref{PvsE} reads:%
\begin{equation}
\mathbf{P}=\left( \mathbb{I}\omega ^{2}-\mathbb{M}\right) ^{-1}\mathbb{R}%
\mathbf{E\equiv }\mathbb{\chi }\mathbf{E,}  \label{Matrix Eq}
\end{equation}%
where $\mathbb{R}\ $is a diagonal matrix with $R_{i,i}=-(4/3)\pi
r_{i}^{3}\omega _{\mathrm{P}i}^{2}\epsilon _{0}$ and $\mathbb{M}$ is\ a
tridiagonal matrix, with

\begin{equation}
M_{0,0}\equiv \widetilde{\omega }_{\mathrm{SP}0}^{2}=\omega _{\mathrm{SP}%
0}^{2}-\mathrm{i}\eta _{_{0}}\omega
\end{equation}%
for the LE-NP, while for any of the equidistant identical NPs along the
chain
\begin{equation}
M_{i,i}\equiv \widetilde{\omega }_{\mathrm{SP}}^{2}=\omega _{\mathrm{SP}%
}^{2}-\mathrm{i}\eta \omega ,~(\forall i\neq 0).
\end{equation}%
The dipole-dipole coupling strengths are
\begin{equation}
M_{i,j}\equiv \omega _{\mathrm{X}i,j}^{2}=\frac{\gamma ^{_{T,L}}\omega _{%
\mathrm{P}i}^{2}}{3n^{2}}\left( \frac{r_{i}}{d_{i,j}}\right) ^{3},
\end{equation}%
with $\gamma ^{T}=1$, and $\gamma ^{L}=-2$.\ Since we are interested in the
particular case where only the LE-NP is different from the rest, it is
convenient adopt the nearest-neighbor coupling as $\omega _{\mathrm{X}%
i,j}^{2}=\omega _{\mathrm{X}}^{2}$\ for $i$\ and $j\neq 0$.

This description is accurate for: 1) $kd\ll 1$ and 2) $r/d\lessapprox 1/3$,
where higher order multipoles are negligible \cite{Higher-Multi}. These
conditions require small $r$'s and hence a negligible radiation damping
correction \cite{CDA}.

Clearly, $\mathbb{\chi }$ is a response function (RF), hence excitation
dynamics between the different sites $i$ and $j$ is determined by the
corresponding matrix elements of $\mathbb{\chi }$. In this way, the square
dipole moment of the $m^{\mathrm{th}}$\ NP, $|P_{m}|^{2}$, produced when the
LE-NP\ is externally excited with an electric field $E_{0}$, is:%
\begin{equation}
|P_{m}|^{2}=|\chi _{_{m0}}|^{2}|E_{0}|^{2}.  \label{D^2}
\end{equation}%
Notice that $\left( \mathbb{I}\omega ^{2}-\mathbb{M}\right) ^{-1}$can be
identified with a Green's function \cite{GF,GFsimple}. The precise
correspondence is presented in the Appendix and serves to exploit the
analytical tools developed in this context.

\section{III. POLE ANALYSIS OF THE RESPONSE FUNCTION}

Since $\mathbb{M}$\ is tridiagonal, $\chi _{_{ij}}\left( \omega ^{2}\right) $%
\ admits exact analytical expressions as continued fractions \cite%
{GFsimple,Thouless}. For finite systems, $\chi _{ij}\left( \omega
^{2}\right) $ has\ a set of isolated poles at the eigenvalues of $\mathbb{M}$
whose real and imaginary parts are respectively the eigenfrequencies and
their damping\textit{.} Extending $\mathbb{M}$\ to an infinite case enabled
us to find a close expression for the response of our system at an arbitrary
position $m$\ given an excitation at position $i=0$:%
\begin{equation}
\chi _{_{m0}}(\omega )=\chi _{_{00}}(\omega )\alpha _{1,0}^{1/2}e^{-m/\xi
(\omega )},  \label{DN0}
\end{equation}%
where
\begin{equation}
\alpha _{1,0}=\omega _{\mathrm{X}1,0}^{4}/\omega _{\mathrm{X}}^{4},
\end{equation}%
accounts for the surface asymmetry, while\
\begin{equation}
\xi ^{-1}(\omega )=\ln (\omega _{\mathrm{X}}^{2}/\Pi )=\kappa \pm ik
\end{equation}%
\ \ is a generalized wave vector. The term $\chi _{_{00}}$ is the RF of the
LE-NP,%
\begin{equation}
\chi _{_{00}}=\frac{R_{00}}{\left[ \omega ^{2}-\widetilde{\omega }_{\mathrm{%
SP}0}^{2}\right] -\alpha \Pi (\omega )}.  \label{D00}
\end{equation}%
where the factor \
\begin{equation}
\alpha =\sqrt{\alpha _{1,0}\alpha _{0,1}}
\end{equation}%
\ accounts for the effective coupling strengths, \textit{i.e.} $\alpha =0$\
describes an isolated LE-NP. For small $\alpha $, the peak at $\widetilde{%
\omega }_{\mathrm{SP}0}^{2}$ is further shifted and broadened by $\alpha \Pi
(\omega ),$ a complex \textquotedblleft self energy\textquotedblright\
accounting for the linear array. When $N\rightarrow \infty $:
\begin{align}
\Pi (\omega )& =\tfrac{1}{2}\left[ \omega ^{2}-\widetilde{\omega }_{\mathrm{%
SP}}^{2}\right] -  \label{Eq-Pi} \\
& \mathrm{sgn}(\omega ^{2}-\omega _{\mathrm{SP}}^{2})\tfrac{1}{2}\sqrt{\left[
\omega ^{2}-\widetilde{\omega }_{\mathrm{SP}}^{2}\right] ^{2}-4\omega _{%
\mathrm{X}}^{4}}.  \notag
\end{align}%
In the weak damping limit (WDL), \textit{i.e.} $\eta \rightarrow 0^{+}$, the
propagating frequencies are given by the usual dispersion relation \cite%
{Brongersma},
\begin{equation}
\omega ^{2}(k)=\ \omega _{\mathrm{SP}}^{2}-2\omega _{\mathrm{X}}^{2}\cos
(kd),
\end{equation}%
with wave number $k$ $\in \lbrack -\pi /d,\pi /d]$. Within the
\textquotedblleft passband\textquotedblright\ $|\omega ^{2}-\omega _{\mathrm{%
SP}}^{2}|$\ $\leq 2\omega _{\mathrm{X}}^{2}$, each frequency component of
the excitation propagates with group velocity
\begin{equation}
v_{g}=\frac{d}{2\omega }\sqrt{4\omega _{\mathrm{X}}^{4}-\left[ \omega
^{2}-\omega _{\mathrm{SP}}^{2}\right] ^{2}}.
\end{equation}%
\ Components outside the passband decay exponentially along the chain within
the localization length $\kappa ^{-1}$. The inclusion of electronic damping $%
\eta $ adds a further decay and smears out the dispersion relation. The
overall behavior of Eq.\ref{DN0} is consistent with the numerical solutions
including full retardation effects under similar conditions \cite{Sarychev}.

The local density of plasmonic states (LDPS) at site $i=0$ is given by $%
\mathrm{Im}(\chi _{_{00}})$. In the WDL, LDPS quantifies the participation
of this site on the different eigenfrequencies in a range $\mathrm{d}\omega $
around $\omega .$ More generally, for finite $\eta $, $\omega \mathrm{Im}%
(\chi _{_{00}})$ is proportional to the power absorbed when site $0$ is
irradiated with frequency $\omega $.

The case $\alpha =0$ of Eq. \ref{D00} exemplifies the general behavior\ of
finite systems, where poles of the RP (zeros in the denominator of $\chi
_{_{00}})$\ determine the frequencies of maximum energy absorption. In this
situation, dissipation occurs due to the damping processes accounted by $%
\eta $. In an infinite system, a new mechanism appears as $\mathrm{Im}(\Pi )$
also describes the irreversible energy spread through the chain. In this
case, Eq. \ref{D00} has \textquotedblleft poles\textquotedblright , $\omega
_{pole}$, which solve:%
\begin{align}
\lbrack \omega _{pole}^{2}-\omega _{\mathrm{SP}0}^{2}-\frac{\alpha }{2}%
(\omega _{pole}^{2}-\omega _{\mathrm{SP}}^{2})]^{2}& =  \label{poles} \\
\frac{\alpha ^{2}}{4}[(\omega _{pole}^{2}-\omega _{\mathrm{SP}%
}^{2})^{2}-4\omega _{\mathrm{X}}^{4}].  \notag
\end{align}%
The solution of this equation is:%
\begin{align}
\widetilde{\omega }_{pole}^{2}& =\frac{\left[ \beta -\alpha \left( \beta
+1\right) /2\right] }{\left( 1-\alpha \right) }\pm  \label{Eq-Poles} \\
& \frac{\alpha }{2(1-\alpha )}\sqrt{(1-\beta )^{2}-4V^{2}(1-\alpha )},
\notag
\end{align}%
where $\widetilde{\omega }_{pole}^{2}=\frac{\omega _{pole}^{2}}{\widetilde{%
\omega }_{\mathrm{SP}}^{2}}$, $\beta =\frac{\widetilde{\omega }_{\mathrm{SP}%
0}^{2}}{\widetilde{\omega }_{\mathrm{SP}}^{2}}$, and $V=\frac{\omega _{%
\mathrm{X}}^{2}}{\widetilde{\omega }_{\mathrm{SP}}^{2}}$. The analysis of
the different types of \textquotedblleft poles\textquotedblright\ or
solutions of Eq. \ref{Eq-Poles} is simplify in the WDL. In this case, two
types of non-physical poles appear: 1) The pole has a positive imaginary
part. 2) The pole is real but it corresponds to a divergence of the original
equation ($\mathbb{\chi }_{_{00}}$) with a non-physical self-energy ($%
\mathrm{Im}(\Pi )<0$). This second alternative can appear because, in order
to obtain a closed expression for $\omega _{pole}$, the denominator of $%
\mathbb{\chi }_{_{00}}$ should be squared (Eq. \ref{poles}). This makes
physical and non-physical self-energies indistinguishable. Therefore, both
solutions are present in Eq. \ref{Eq-Poles}. The second type of non-physical
poles are associated with \textit{virtual} states, because even though they
do not appear as resonances, they affect the LDPS within the passband, in
the same way as \textit{localized} or "real" states.

\begin{figure}[th]
\begin{center}
\includegraphics[trim=0.5in 0.3in 0.3in 0.0in, width=2.9in]{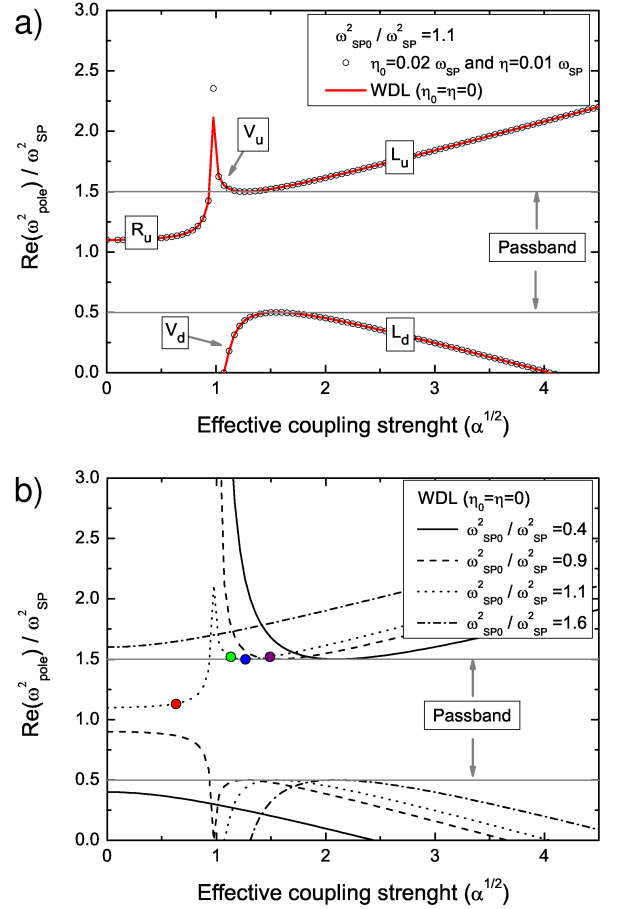}
\end{center}
\caption{(Color online) Real part of poles of $\protect\chi $ as function of
$\protect\alpha ^{1/2}$. Here, $\protect\omega _{\mathrm{X}}^{2}/\protect%
\omega _{\mathrm{SP}}^{2}=0.25$. In a) three different regimes are shown:
discrete localized levels (L$_{d}$ and L$_{u}),$ resonant level (R$_{u})$
and virtual states (V$_{d}$ and V$_{u})$. Subscript \textbf{u} and \textbf{d}
stand for \textquotedblleft up\textquotedblright\ and \textquotedblleft
down\textquotedblright , the possible positions of $\mathrm{Re}(\protect%
\omega _{pole}^{2})$ relative to $\protect\omega _{\mathrm{SP}}^{2}$. The
numerical solutions for finite $\protect\eta $ are compared with the
analytical WDL. In panel b) we explore the dependence of the poles on $%
\protect\omega _{\mathrm{SP}0}$. The colored dots indicate the parameters
used in the discussion of the LDPS (see Fig. 3).}
\label{Fig-polos}
\end{figure}

Fig. \ref{Fig-polos} shows the real part of the poles of the response
function, $\omega _{pole}$, as function of $\alpha ^{1/2}$ for four
different values of $\omega _{\mathrm{SP}0}$. The figure also compare the
case of the analytical solution resulting from the WDL of Eq. \ref{Eq-Poles}
with its numerical evaluation for a case with realistic damping. Numerical
evaluation is required for $\eta \neq 0$ because $\beta $, $V$ and $%
\widetilde{\omega }_{pole}^{2}$ depend on $\omega $ in this case. Notice the
accuracy of the WDL approximation. These poles provide\ for discrete \textit{%
localized} levels (L), \textit{resonant} levels (R) and \textit{virtual}
states (V) according to the parametric region.

\begin{figure}[th]
\begin{center}
\includegraphics[trim=0.5in 0.3in 0.3in 0.0in, width=2.9in]{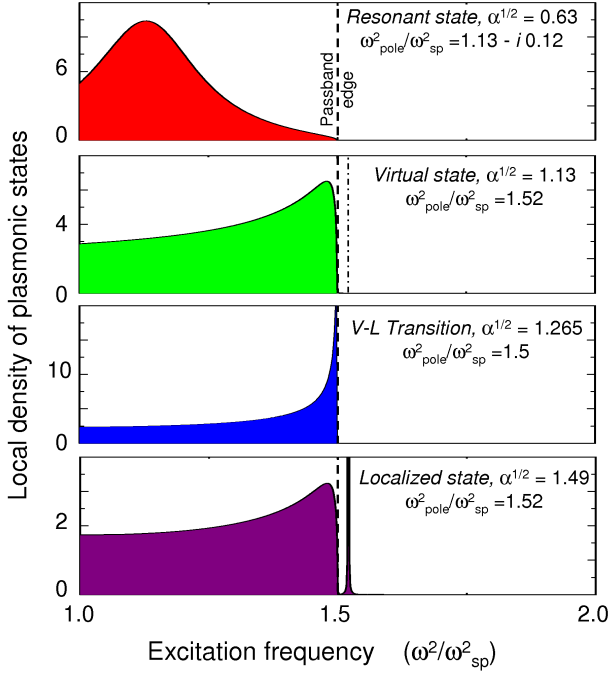}
\end{center}
\caption{(Color online) LDPS as function of the excitation frequency for
different $\protect\alpha $ values. Here, $\protect\omega _{\mathrm{SP}%
0}^{2}/\protect\omega _{\mathrm{SP}}^{2}=1.1$ and $\protect\omega _{\mathrm{X%
}}^{2}/\protect\omega _{\mathrm{SP}}^{2}=0.25$. Also, $\protect\eta =\protect%
\eta _{0}=0$ except for the panel showing a localized state case, where $%
\protect\eta =10^{-5}$. Notice the identical shape of the LDPS near the
passband edge for both \textit{virtual} and \textit{localized} states, and
the divergence at the \textit{virtual-localized} transition.}
\label{LDPS}
\end{figure}

In the WDL, when a pole is complex with a negative imaginary part, its real
part lies within the passband and corresponds to the eigenfrequency of the
LE-NP (Fig. \ref{LDPS}), while its imaginary part roughly represents the
decay rate. This is the case of a \textit{resonant} state. When the pole is
real, the usual situation is that the system has a \textit{localized}
eigenmode whose eigenfrequency lies outside the passband (Fig. \ref{LDPS}).
An excitation at this frequency will remain indefinitely within the
localization length $1/\kappa $. These two situations would typically
exhaust the analysis. However, for quantum systems it has recently become
clear that the transition between these two regimes, although covering a
very narrow parametric range, has subtle and unique properties: there is a
real \textquotedblleft pole\textquotedblright\ which nevertheless does not
correspond to an eigenstate of the system (see \cite{Dente,Virtual}). As
such, one might not know what to expect. This situation worsens in a
plasmonic case where such parametric region broadens. Even though, they are
not physical poles, in fact they do not solve $[\omega ^{2}-\widetilde{%
\omega }_{\mathrm{SP}0}^{2}]-\alpha \Pi =0$, they still affect the LDPS, and
hence the RF, within the passband (Fig. \ref{LDPS}) in a similar way than
\textit{localized} states. In both cases, the LDPS is modulated at the band
edges by a\ divergent factor $1/(\omega ^{2}-\omega _{pole}^{2})$.

This relationship between the poles of RF and the LDPS of the LE-NP,\ can be
explicitly written in the WDL:

\begin{equation}
\mathrm{LDPS}=\dfrac{R_{00}}{\omega _{\mathrm{SP}}^{2}}\times N\left( \omega
\right) \times c\times L\left( \omega \right) ,  \label{Im(D00WDL)}
\end{equation}%
where $\widetilde{\omega }^{2}=\omega ^{2}/\omega _{\mathrm{SP}}^{2}$, $c$
is a normalization constant
\begin{equation}
c=\frac{2V^{2}}{\sqrt{4V^{2}(1-\alpha )-(1-\beta )^{2}}},  \label{C}
\end{equation}%
$L$ is a Lorentzian function (in a plot as function of $\widetilde{\omega }%
^{2}$),%
\begin{equation}
L\left( \omega \right) =\frac{\widetilde{\Gamma }^{2}}{\left( \widetilde{%
\omega }^{2}-\widetilde{\omega }_{0}^{2}\right) ^{2}+\widetilde{\Gamma }^{4}}%
,  \label{L}
\end{equation}%
and $N\left( \omega \right) $ is proportional to the LDPS at the surface
site of a semi-infinite chain of identical NPs,%
\begin{equation}
N\left( \omega \right) =\frac{\sqrt{4V^{2}-\left( 1-\widetilde{\omega }%
^{2}\right) ^{2}}}{2V^{2}}.  \label{N}
\end{equation}%
In the \textit{resonant} state regime, the maximum and the width of $L\left(
\omega \right) $ are given by the real and imaginary part of $\widetilde{%
\omega }_{pole}^{2}$,
\begin{equation}
\widetilde{\omega }_{0}^{2}=\left( \frac{\omega _{0}}{\omega _{\mathrm{SP}%
}^{{}}}\right) ^{2}=\frac{\left[ \beta -\alpha \left( \beta +1\right) /2%
\right] }{\left( 1-\alpha \right) }  \label{Repolo}
\end{equation}%
and
\begin{equation}
\widetilde{\Gamma }^{2}=\left( \frac{\Gamma }{\omega _{\mathrm{SP}}^{{}}}%
\right) ^{2}=\frac{\alpha }{2(1-\alpha )}\sqrt{4V^{2}(1-\alpha )-(1-\beta
)^{2}}.  \label{Impolos}
\end{equation}%
In the \textit{virtual} and \textit{localized} state regimes, these
expressions are still valid but $L\left( \omega \right) $ is no longer a
Lorentzian function as $\widetilde{\Gamma }^{2}$ is imaginary. In this case,
it is better to write the LDPS at the LE-NP as:

\begin{align}
\mathrm{LDPS}=& \dfrac{R_{00}}{\omega _{\mathrm{SP}}^{2}}\times N\left(
\omega \right)   \label{divergences} \\
& \times \frac{\alpha V^{2}}{(1-\alpha )}\frac{1}{(\widetilde{\omega }^{2}-%
\widetilde{\omega }_{pole+}^{2})(\widetilde{\omega }^{2}-\widetilde{\omega }%
_{pole-}^{2})},  \notag
\end{align}

Notice the divergences that appear at the \textit{virtual-localized}
transition, \textit{i.e.} when $\omega _{pole}$ reaches the passband edge.
These divergences will strongly favor the excitation transfer. Fig.\ref%
{Ejemplos-RVL} illustrates the frequency dependent excitation transfer along
the chain for two cases: a) a system with a \textit{resonant\ }state and b)
a system with both, a \textit{virtual} and a \textit{localized} state.
\begin{figure}[th]
\begin{center}
\includegraphics[trim=0.5in 0.3in 0.3in 0.0in, width=2.9in]{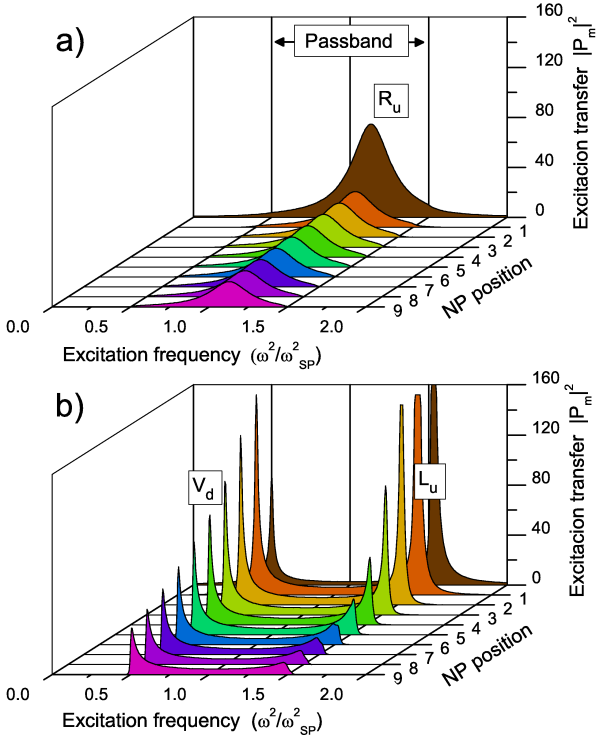}
\end{center}
\caption{(Color online) Square dipolar moment (in arbitrary units) as
function of the NP position and excitation frequency. Here, $\protect\eta %
=0.01\protect\omega _{\mathrm{SP}}$, $\protect\eta _{0}=0.02\protect\omega _{%
\mathrm{SP}}$, $\protect\omega _{\mathrm{SP}0}^{2}/\protect\omega _{\mathrm{%
SP}}^{2}=1.1$, and $\protect\omega _{\mathrm{X}}^{2}/\protect\omega _{%
\mathrm{SP}}^{2}=0.25$. In \textbf{Fig.a }$\protect\alpha ^{1/2}=0.63$ and
in \textbf{Fig.b} $\protect\alpha ^{1/2}=1.55$. \textbf{L}, \textbf{V}, and
\textbf{R} stand for \textit{localized}, \textit{virtual} and \textit{%
resonant} states respectively, while \textbf{u} and \textbf{d} stand for up
and down, the position of the pole relative to $\protect\omega _{\mathrm{SP}%
} $.}
\label{Ejemplos-RVL}
\end{figure}

As occurs for the oscillating-overdamped transition, and many of the
spectral bifurcations omnipresent in non-Hermitian Hamiltonians \cite{Rotter}%
, the critical parameters of Eq. \ref{Eq-Poles} occur when the argument
under a square root vanishes and they identify the phase transition in the
dynamical behavior \cite{QDPT}. The transition \textit{resonant-virtual}
occurs for $\mathrm{Im}(\omega _{pole})=0$:%
\begin{equation}
\alpha _{c1}=\frac{2\beta +4V^{2}-\beta ^{2}-1}{4V^{2}},  \label{alfaC1}
\end{equation}%
and the transition \textit{virtual-localized} takes place when $\omega
_{pole}\in \Re $ and $\widetilde{\omega }_{pole}^{2}=1\pm 2V$:%
\begin{equation}
\alpha _{c2(\pm )}=2\pm \frac{(1-\beta )}{V}.  \label{alfaC2}
\end{equation}%
These expressions, obtained in the WDL, will result essential to asses the
role of the poles\ in the excitation transfer.

\section{IV. OPTIMAL\ EXCITATION TRANSFER}

\begin{figure}[th]
\begin{center}
\includegraphics[trim=0.5in 0.3in 0.3in 0.0in, width=3.0in]{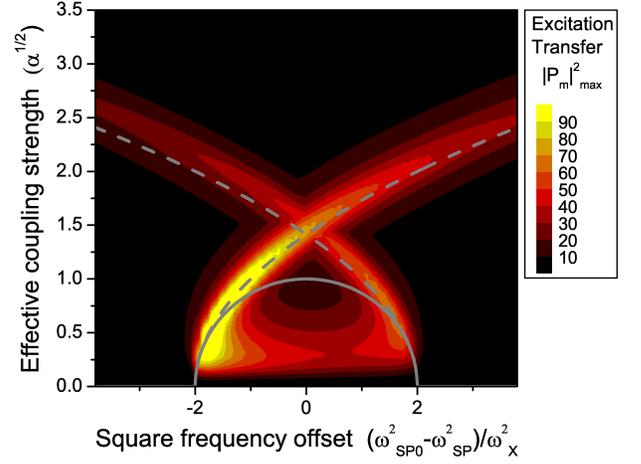}
\end{center}
\caption{(Color online) Color scale shows the maximum excitation transfer to
the 5$^{th}$ NP ($\max |P_{5}(\protect\omega )|^{2}$ in arbitrary units) for
$\protect\eta =0.01\protect\omega _{\mathrm{SP}}$ and $\protect\eta _{0}=0.02%
\protect\omega _{\mathrm{SP}}$. Continuous line delimits the region of
\textit{resonant} states (Eq.\protect\ref{alfaC1}) and dash lines mark the
\textit{virtual-localized} transitions (Eq.\protect\ref{alfaC2}), both
evaluated in the WDL.}
\label{Dfase}
\end{figure}

Fig \ref{Dfase} shows the maximum excitation transfer, $|P_{m}|^{2},$
enabled by a variation of $\omega $ \ at each system configuration.
Superposed are the critical values resulting from Eqs. \ref{alfaC1} and \ref%
{alfaC2}. Consistently with the above discussion, an appreciable transfer
occurs in the \textit{resonant} state regime. However, the maximum appears
at the transition between \textit{virtual} and \textit{localized} states.
Note that the optimal configuration for excitation transfer does not occur
for $\omega _{\mathrm{SP}0}=\omega _{\mathrm{SP}}$ and $\alpha =1,$ where
the LE-NP is indistinguishable from the others, as one might na\"{\i}vely
expect. Instead, it occurs for the highly asymmetric configuration where a
\textit{virtual}-\textit{localized} transition appears.\ In this case, an
excitation of a local mode that is coupled to collective excitations at the
band edge, has components with very slow group velocity. In consequence,
under a continuous irradiation, the excitation can build up. This dynamical
interpretation is consistent with the new strategy developed in the context
of photonic crystals. There, waveguides with slow group velocities are used
as a way of buffering light\cite{Baba-photonic}.\ Alternatively, one sees
that just at the \textit{virtual}-\textit{localized} transition, the
spectroscopic Eq. \ref{DN0} favorably combines its three factors: the
response function on the excited nanoparticle $\chi _{_{00}}(\omega )$\ has
a high intensity peak, this peak is inside the passband, and the relative
effective coupling $\alpha _{1,0}^{1/2}$\ is strong.\textit{\ }
Notice that a high transfer efficiency could be achieved even for $\omega _{%
\mathrm{SP}0}$ so different from $\omega _{\mathrm{SP}}$ that the system
could never form a \textit{resonant} state just by changing $\alpha $. In
analogy with the addatom in an Anderson-Newns model \cite{Dente}, a strong
interaction with the substrate captures a state from the continuum spectrum
to build a second \textit{localized} state that would constitute the
\textquotedblleft antibonding\textquotedblright\ orbital of a dimer \cite%
{Nordlander}. This occurs through a \textit{virtual}-\textit{localized}
transition and thus leads to the optimal excitation transfer shown in Fig %
\ref{Dfase}. Experimentally, these critical points could be achieved by\
properly tuning the distance, radius, shape and material of the NPs.
Additionally, this configuration acts as a very narrow filter for the
external frequency in resonance with the passband edge (See Fig. \ref%
{Ejemplos-RVL}-b).The control of this critical phenomenon opens up many
possibilities for applications. For example, the extreme sensibility of
excitation transfer on $d_{0,1}$ when the system is close to a critical
transition, would enable a new form of plasmon ruler suitable for biological
and chemical applications \cite{Alivisatos}. This occurs because $\alpha $
varies with $d_{0,1}^{-6}$ and, depending on $\omega _{\mathrm{SP}0}$, $%
\omega _{\mathrm{SP}}$ and $\omega _{\mathrm{X}}^{2}$ which give the
frequency offset of Fig .\ref{Dfase}, the system response sweeps through
different regimes within a narrow interval of $\alpha $. Similarly, as small
changes on the refractive index modify dramatically the coupling $\omega _{%
\mathrm{X}}^{2}$ and hence the passband, the excitation transfer will also
be extremely sensitive to the dielectric environment in a system tuned with
the \textit{virtual}-\textit{localized} transition.

\section{V. CONCLUSIONS}

We have demonstrated that, contrary to common wisdom, the highest excitation
transfer does not occur for a system with a well defined resonance but when
a \textit{virtual} state is transformed into a \textit{localized} collective
plasmonic mode whose eigenfrequency is just at the passband edge. The slow
group velocity of an excitation with this critical frequency enables the
excitation buffering and hence favors a strong signal inside the chain. The
extreme sensitivity of this transition to the waveguide and LE-NP parameters
would provide new tools for plasmonics. As the basic model is quite general,
our conclusions are universal in nature and apply to any of the broad class
of systems that can be mapped to a linear array of damped oscillators \cite%
{Examp-Appl}.

\section{ACKNOWLEDGEMENTS}

The authors acknowledge the financial support from CONICET, SeCyT-UNC,
ANPCyT, and MinCyT-C\'{o}rdoba. Discussions with Axel Dente and Hern\'{a}n
Calvo are greatly acknowledged.

\section{APPENDIX. Equivalence between plasmonic and quantum mechanical
magnitudes.}

The correspondence between magnitudes in different models is analogous to
the known Bloch discussion of\ unimpeded electronic motion in crystalline
metals in terms of a periodic 1-d array of coupled pendula as described in
Section 5.2 of ref. [11] and other textbooks. This is summarized in the
following Table.
\begin{widetext}
\begin{equation}
\begin{tabular}{|c|c|}
\hline
$%
\begin{array}{c}
\text{{\Large Plasmonics}} \\
\text{(\textit{Near Field Approximation }} \\
\text{\textit{in the weak damping limit, WDL})}%
\end{array}%
$ & $%
\begin{array}{c}
\text{{\Large Quantum Mechanics}} \\
\text{(\textit{Tight Binding Model})}%
\end{array}%
$ \\ \hline
$P_{i}$ : dipole moment at $i^{\mathrm{th}}$ NP & $c_{i}$ : component of
wavefunction at site $i^{\mathrm{th}}$ \\ \hline
$\mathbb{M}$ : dynamical matrix & $\mathbb{H}$ : Hamiltonian matrix \\ \hline
$\omega _{\mathrm{SP}i}^{2}$ $\equiv M_{i,i}$:$%
\begin{array}{c}
\text{square of surface plasmon } \\
\text{frequency of the }i^{\mathrm{th}}\text{ uncoupled NP}%
\end{array}%
$ & $E_{i}$ $\equiv H_{i,i}$: isolated $i^{\mathrm{th}}$ site energy \\
\hline
$\omega ^{2}$ : excitation frequency & $\varepsilon $ : propagation energy
\\ \hline
$D_{ii}^{(0)}=(\omega ^{2}-\omega _{\mathrm{SP}i}^{2})^{-1}$:$%
\begin{array}{c}
\text{decoupled plasmonic} \\
\text{Green's function (GF)}%
\end{array}%
$ & $G_{ii}^{(0)}=(\varepsilon -E_{i})^{-1}$:$%
\begin{array}{c}
\text{ local Green's function (GF)} \\
\text{(locator)}%
\end{array}%
$ \\ \hline
$\omega _{\mathrm{X}i,j}^{2}$ $\equiv M_{i,j}$: dipole-dipole coupling
strength & $V_{i,j}$ $\equiv H_{i,j}$: hopping amplitude \\ \hline
$\mathbb{D}=(\omega ^{2}\mathbb{I}-\mathbb{M})^{-1}$ : plasmonic GF & $%
\mathbb{G}=(\varepsilon \mathbb{I}-\mathbb{H})^{-1}$ : GF \\ \hline
$D_{00}^{{}}=(\omega ^{2}-\omega _{\mathrm{SP}0}^{2}-\alpha \Pi )^{-1}$ :
surface site GF & $G_{00}^{{}}=(\varepsilon -E_{0}-\alpha \Sigma )^{-1}$:
surface site GF \\ \hline
$\Pi $ : $%
\begin{array}{c}
\text{plasmonic waveguide's} \\
\text{self energy}%
\end{array}%
$ & $\Sigma $ : $%
\begin{array}{c}
\text{linear chain's} \\
\text{self energy}%
\end{array}%
$ \\ \hline
$%
\begin{array}{c}
\Pi =\dfrac{\omega _{\mathrm{X}}^{4}}{\omega ^{2}-\omega _{\mathrm{SP1}}^{2}-%
\dfrac{\omega _{\mathrm{X}}^{4}}{\omega ^{2}-\omega _{\mathrm{SP2}}^{2}-%
\dfrac{\omega _{\mathrm{X}}^{4}}{\cdots {\cdots -\dfrac{\omega _{%
\mathrm{X}}^{4}}{\omega ^{2}-\omega _{\mathrm{SP}\text{\textrm{N}}}^{2}}}}}}
\\
\text{:finite system}%
\end{array}%
$ & $%
\begin{array}{c}
\Sigma =\dfrac{V^{2}}{\varepsilon -E_{1}-\dfrac{V^{2}}{\varepsilon -E_{2}-%
\dfrac{V^{2}}{\cdots {\cdots -\dfrac{V^{2}}{\varepsilon -E_{N}}}}}%
} \\
\text{:finite system}%
\end{array}%
$ \\ \hline
$%
\begin{array}{c}
\Pi =\dfrac{\omega _{\mathrm{X}}^{4}}{\omega ^{2}-\omega _{\mathrm{SP}%
}^{2}-\Pi } \\
:\text{infinite system}%
\end{array}%
$ & $%
\begin{array}{c}
\Sigma =\dfrac{V^{2}}{\varepsilon -E-\Sigma } \\
:\text{infinite system}%
\end{array}%
$ \\ \hline
\end{tabular}
\
\end{equation}
\end{widetext}

\end{document}